\documentclass{aastex}

\usepackage{spr-astr-addons}
\usepackage{graphicx,amssymb}
\usepackage[normalem]{ulem}

\shorttitle{Dynamo and Winds}
\shortauthors{Beck}

\begin{document}

\title{Galactic Dynamos and Galactic Winds}

\author{Rainer Beck}
\affil{Max-Planck-Institut f\"ur Radioastronomie, Auf dem H\"ugel
69, 53121 Bonn, Germany}

\begin{abstract}
Spiral galaxies host dynamically important magnetic fields which can
affect gas flows in the disks and halos. Total magnetic fields in
spiral galaxies are strongest (up to $30~\mu$G) in the spiral arms
where they are mostly turbulent or tangled. Polarized synchrotron
emission shows that the resolved regular fields are generally
strongest in the interarm regions (up to $15~\mu$G). Faraday
rotation measures of radio polarization vectors in the disks of
several spiral galaxies reveal large-scale patterns which are
signatures of coherent fields generated by a mean-field dynamo. --
Magnetic fields are also observed in radio halos around edge-on
galaxies at heights of a few kpc above the disk. Cosmic-ray driven
galactic winds transport gas and magnetic fields from the disk into
the halo. The halo scale height and the electron lifetime allow to
estimate the wind speed. The magnetic energy density is larger than
the thermal energy density, but smaller than the kinetic energy
density of the outflow. There is no observation yet of a halo with a
large-scale coherent dynamo pattern. A global wind outflow may
prevent the operation of a dynamo in the halo. -- Halo regions with
high degrees of radio polarization at very large distances from the
disk are excellent tracers of interaction between galaxies or ram
pressure of the intergalactic medium. The observed extent of radio
halos is limited by energy losses of the cosmic-ray electrons. --
Future low-frequency radio telescopes like LOFAR and the SKA will
allow to trace halo outflows and their interaction with the
intergalactic medium to much larger distances.
\end{abstract}


\section{Introduction}

Galactic halos are dynamical systems receiving their energy from the
underlying star-forming disk. Hot thermal gas from supernova
remnants or superbubbles can drive fountain outflows, but no steady
wind \citep{breit91}. Cosmic rays accelerated in supernova remnants
can provide the pressure to drive a wind. Cosmic rays also drive
buoyant loops of magnetic fields via the Parker instability
\citep{hanasz02} (see also Avillez, this volume) and a fast dynamo
which amplifies the magnetic field \citep{parker92,moss99,hanasz04}.
Streaming instabilities of the cosmic-ray flow excite plasma waves
in the halo and couple the gas to the outflow \citep{breit91}.
Magnetic reconnection can heat the gas \citep{zimmer97}. Outflows
from starburst galaxies in the early Universe may have magnetized
the intergalactic medium \citep{kronberg99}, but the role of
present-day galaxies is still unclear. Understanding the interaction
between the gas and the magnetic field in the outflow is the key to
understand the physics of halos and their role during the evolution
of galaxies.

\section{The tools of radio synchrotron emission}

The intensity of synchrotron emission is a measure of the number
density of cosmic-ray electrons in the relevant energy range and of
the strength of the total magnetic field component in the sky plane.
The degree of linear polarization can be up to 75\%. Any variation
of the field orientation within the beam reduces the degree of
polarization. Polarized emission emerges from regular fields, which
are fields having a constant direction within the telescope beam, or
from anisotropic magnetic fields, which are generated from random
magnetic fields by a compressing or shearing gas flow and frequently
reverse their direction by 180\degr\ on scales smaller than the
telescope beam. Unpolarized synchrotron emission indicates fields
with random directions which have been tangled or created by
turbulent gas flows.

At short radio wavelengths the orientation of the observed
polarization vector is perpendicular to the field orientation. The
orientation of the polarization vectors is changed in a magnetized
thermal plasma by Faraday rotation. The rotation angle increases
with the plasma density, the strength of the component of the field
along the line of sight and the square of the observation
wavelength. As the rotation angle is sensitive to the sign of the
field direction, only regular fields can give rise to Faraday
rotation, while anisotropic and random fields do not. For typical
plasma densities and regular field strengths in the interstellar
medium of galaxies, Faraday rotation becomes significant at
wavelengths larger than a few centimeters. Measurements of the
Faraday rotation from multi-wavelength observations allow to
determine the strength and direction of the regular field component
along the line of sight. Its combination with the total intensity
and the polarization vectors can yield the three-dimensional picture
of the magnetic field and allows to distinguish the three field
components: regular, anisotropic and random.

\section{Magnetic fields in spiral arms and bars}

The total magnetic field strengths can be determined from the
intensity of the total synchrotron emission, assuming energy
equipartition between the total magnetic field and the total cosmic
rays, and a ratio between the numbers of cosmic-ray protons and
electrons in the relevant energy range (usually $\simeq100$). The
equipartition assumption seems to be valid in galaxies on large
spatial and time scales, but deviations probably occur on local
scales. The typical average equipartition strength of the total
magnetic field in spiral galaxies is about $10~\mu$G. Radio-faint
galaxies like M~31 and M~33, our Milky Way's neighbors, have weaker
total magnetic fields (about $5~\mu$G), while gas-rich galaxies with
high star-formation rates, like M~51, M~83 and NGC~6946, have
average total field strengths of $15~\mu$G. The mean energy density
of the magnetic field and of the cosmic rays in NGC~6946 is $\simeq
10^{-11}$~erg~cm$^{-3}$, about 10 times larger than that of the
ionized gas, but similar to that of the turbulent gas motions across
the whole star-forming disk \citep{beck07a}. The magnetic energy may
even dominate in the outer disk. The strongest fields
($50-100~\mu$G) are found in starburst galaxies, like M~82 and the
``Antennae'' NGC~4038/9, and in nuclear starburst regions, like in
the centers of NGC~1097 and other barred galaxies \citep{beck05}.

Spiral arms in total radio emission appear very similar to those
observed in the far-infrared. The total equipartition field strength
in the arms can be up to $30~\mu$G. The degree of radio polarization
within the spiral arms is only a few \%; hence the field in the
spiral arms must be mostly tangled or random within the telescope
beam, which typically corresponds to a few 100~pc. Strong random
fields in spiral arms are probably generated by turbulent gas
motions or the turbulent dynamo \citep{brand05}. In contrast, the
ordered (regular or anisotropic) fields traced by the polarized
synchrotron emission are generally strongest ($10-15~\mu$G) in the
regions {\em between} the optical spiral arms, oriented parallel to
the adjacent optical spiral arms. In several galaxies the field
forms ``magnetic arms'' between the optical arms, like in NGC~6946
\citep{beck07a}. These are probably generated by the mean-field
dynamo (Sect.~\ref{sect:dynamo}). In galaxies with strong density
waves some of the ordered field is concentrated at the inner edge of
the spiral arms, e.g. in M~51 \citep{patrikeev06}.

The ordered magnetic field forms spiral patterns in almost every
galaxy, even in flocculent and bright irregular ones which lack any
optical spiral structure \citep{beck05}, a strong argument for
dynamo action. Spiral fields are also observed in the central
regions of galaxies and in circum-nuclear gas rings.

In galaxies with massive bars, the field seems to follow the gas
flow. As there (inside the corotation radius) the gas rotates faster
than the spiral or bar pattern of a galaxy, a shock occurs in the
cold gas which has a small sound speed, while the warm, diffuse gas
is only slightly compressed. As the observed compression of the
field in spiral arms and bars is also small, the ordered field is
coupled to the warm gas and is strong enough to affect the flow of
the warm gas \citep{beck+05}. The polarization pattern is an
excellent tracer of the gas flow in the sky plane and hence
complements spectroscopic measurements.

\section{Galactic dynamos}
\label{sect:dynamo}

\begin{figure}
\begin{center}
  \includegraphics[angle=90,width=0.35\textwidth]{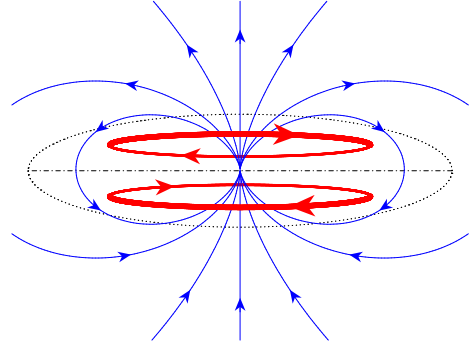}
 \end{center}
\caption{Dynamo field of type A0: The toroidal field component
(thick lines) is axisymmetric in the plane (azimuthal mode $m=0$)
and reverses its sign across the plane. The poloidal field component
(thin lines) is of dipole type \citep{han02}.} \label{fig:A0}
\end{figure}

The origin of the first magnetic fields in the Universe is still a
mystery \citep{widrow02}. Protogalaxies probably were already
magnetic due to field ejection from the first stars or from jets
generated by the first black holes. A large-scale primordial field
in a young galaxy is hard to maintain because the galaxy rotates
differentially, so that field lines get strongly wound up during
galaxy evolution, in contrast to the observations which show large
pitch angles. This calls for a mechanism to sustain and organize the
magnetic field. The most promising mechanism is the dynamo
\citep{beck96,brand05} which generates large-scale regular fields,
even if the seed field was turbulent.

The regular field structure obtained in models of the mean-field
$\alpha\Omega$--dynamo, driven by turbulent gas motions and
differential rotation, is described by modes of different azimuthal
symmetry in the disk and vertical symmetry perpendicular to the disk
plane. Several modes can be excited in the same object.

In spherical bodies like stars, planets or galaxy halos, the
strongest mode is an antisymmetric toroidal field with a sign
reversal across the equatorial plane, surrounded by a poloidal
dipole field which is continuous across the plane (mode A0,
Fig.~\ref{fig:A0}). For thick galaxy disks or halos the S0 and A0
modes of strong dynamos are oscillatory, i.e. the field torus
migrates radially, so that large-scale reversals in the disk at
certain radii are expected \citep{elstner92}.

In flat objects like galaxy disks, the strongest mode (S0) is a
toroidal field which is symmetric with respect to the plane and has
the azimuthal symmetry of an axisymmetric spiral in the plane,
without sign reversals, surrounded by a weaker poloidal field of
quadrupole structure with a reversal of the vertical field component
across the equatorial plane. The next strongest mode is of
bisymmetric spiral shape (S1) with two sign reversals within the
disk, followed by more complicated modes \citep{bary87}.

The standard mean-field $\alpha\Omega$--dynamo in galaxy disks
amplifies the field and builds up large-scale coherent fields within
about $10^9$~yr \citep{beck94b}. Faster amplification is possible
when cosmic-ray driven Parker loops and field reconnection are
involved \citep{parker92, hanasz98, moss99, hanasz04}. The
small-scale or fluctuation dynamo \citep{brand05} amplifies
turbulent, incoherent magnetic fields within $10^6$--$10^7$~yr.

\begin{figure}
\begin{center}
  \includegraphics[width=0.35\textwidth]{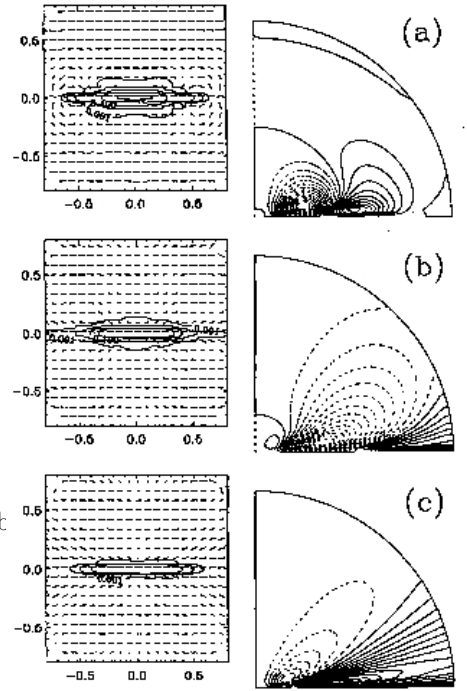}
 \end{center}
\caption{Dynamo model for the galaxy NGC~891 with a galactic wind:
predicted contours of polarized radio emission and vectors (left)
and field geometry (right) for a radial wind with a terminal
velocity of (a) 0, (b) 50 and (c) 200~km/s. The solid and dashed
lines indicate different field directions; the field is oscillatory
and reverses its direction along radius and height
\citep[from][Fig.~9]{brand93}.} \label{fig:wind}
\end{figure}

\begin{figure}
\begin{center}
  \includegraphics[bb = 27 27 492 484,width=0.4\textwidth,clip=]{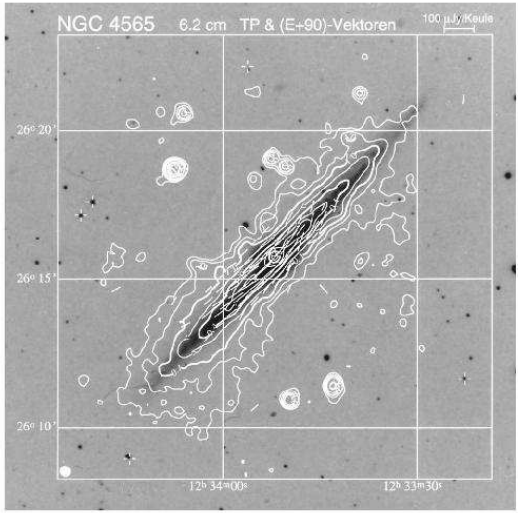}
 \end{center}
\caption{Optical image of the edge-on spiral galaxy NGC~4565,
overlaid by contours of the intensity of the total radio emission at
6~cm wavelength and polarization vectors, combined from observations
with the VLA and the Effelsberg 100-m telescope \citep{dumke97}. The
field lines are mostly parallel to the disk. (Copyright: MPIfR Bonn)
} \label{fig:n4565}
\end{figure}

\begin{figure}
\begin{center}
  \includegraphics[bb = 27 25 486 626,width=0.35\textwidth,clip=]{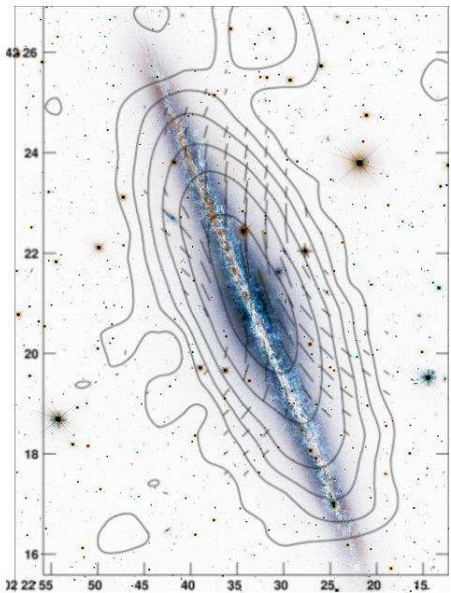}
 \end{center}
\caption{Optical image of the edge-on spiral galaxy NGC~891,
overlaid by contours of the intensity of the total radio emission at
3.6~cm wavelength and polarization vectors, observed with the
Effelsberg 100-m telescope \citep{krause07}. The field lines are
parallel to the disk near the plane, but turn vertically above and
below the disk. (Copyright: MPIfR Bonn) } \label{fig:n891}
\end{figure}

Kinematical dynamo models including the velocity field of a global
galactic wind show field structures which are parallel to the plane
in the inner disk, but open radially outwards, depending on the wind
speed and direction \citep{brand93}. The model for the galaxy
NGC~891 shows an oscillating field (Fig.~\ref{fig:wind}), while for
other galaxy models a slow wind speed of 50~km/s was found to be
sufficient to change the oscillating mode into a steady one.

For fast global winds the advection time for the field may become
smaller than the dynamo amplification time, so that the dynamo
cannot operate and the field gas becomes frozen into the flow.
Non-uniform local (``spiky'') winds allow dynamo action, but modify
the field configuration into oscillating dipoles or quadrupoles with
dominating vertical components and frequent, migrating reversals
\citep{elstner95}. However, strong fields may back-react onto the
wind. Dynamical models are needed, including the interplay between
the gas flow and the magnetic field.

Dynamo modes can be identified from the pattern of polarization
angles and Faraday rotation measures (RM) in multi-wavelength radio
observations of galaxy disks with moderate inclination
\citep{elstner92,krause90} or from RM data of polarized background
sources \citep{stepanov08}. The disks of a few spiral galaxies
indeed reveal large-scale RM patterns, as predicted. The Andromeda
galaxy M~31 hosts a dominating axisymmetric disk field, the basic
azimuthal dynamo mode S0 \citep{fletcher04}, which extends to at
least 15~kpc distance from the center \citep{han98}. Other
candidates for a dominating axisymmetric disk field are the nearby
spiral IC~342 \citep{krause89} and the irregular Large Magellanic
Cloud (LMC) \citep{gaensler05}. The magnetic arms in NGC~6946 can
be described by a superposition of two azimuthal dynamo modes, where
the dynamo wave is phase shifted with respect to the density wave
\citep{beck07a}. However, in many observed galaxy disks no clear
patterns of Faraday rotation were found. Either several dynamo
modes are superimposed and cannot be distinguished with the limited
sensitivity and resolution of present-day telescopes, or no
large-scale dynamo modes exist and most of the ordered fields traced
by the polarization vectors are anisotropic (with frequent
reversals), due to shearing or compressing gas flows.

\section{Radio halos of edge-on galaxies}

Radio halos are observed around the disks of many edge-on galaxies
\citep{hummel91b,irwin99}, but their radio intensity and extent
varies significantly. The radial extent of radio halos increases
with the size of the actively star-forming parts of the galaxy disk
\citep{dahlem06}. The halo luminosity in the radio range correlates
with those in H$\alpha$ and X-rays \citep{tuell06}, although the
detailed halo shapes vary strongly between the different spectral
ranges. These results suggest that star formation in the disk is the
energy source for halo formation. \citet{dahlem95} argue that the
energy input from supernova explosions per surface area in the
projected disk determines the halo size.

\begin{figure}
\begin{center}
  \includegraphics[bb = 49 176 545 649,width=0.4\textwidth,clip=]{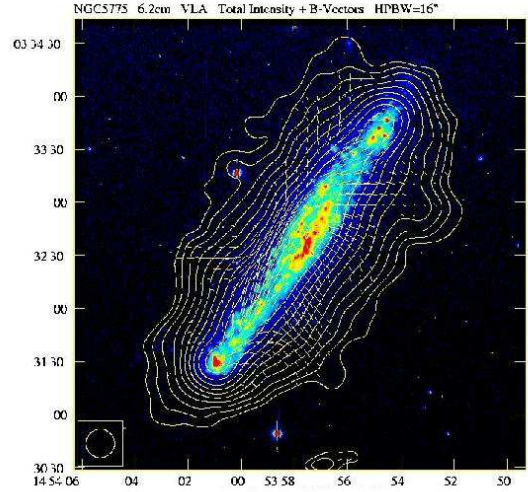}
 \end{center}
\caption{H$\alpha$ image of the edge-on spiral galaxy NGC~5775,
overlaid by contours of the intensity of the total radio emission at
6~cm wavelength and polarization vectors, observed with the VLA
\citep{tuell00}. The field lines are parallel to the disk
near the plane, but become increasingly vertical away from the
plane. (Copyright: Cracow Observatory) } \label{fig:n5775}
\end{figure}

\begin{figure}
\begin{center}
  \includegraphics[width=0.27\textwidth]{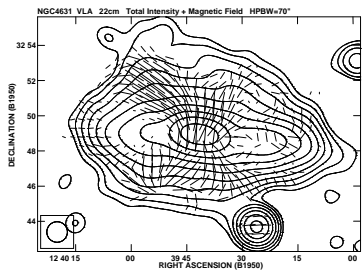}
 \end{center}
\caption{Total radio emission at 22~cm wavelength of the edge-on
spiral galaxy NGC~4631, observed with the VLA \citep{beck05}. The
polarization vectors are corrected for Faraday rotation. The lack of
polarized emission near the plane is due to Faraday depolarization.
(Copyright: MPIfR Bonn) } \label{fig:n4631cm22}
\end{figure}

In spite of their largely different intensities and extents, the
scale heights of radio halos observed at 5~GHz are $\simeq1.8$~kpc
\citep{dumke98,krause04} and surprisingly similar. Their sample of
galaxies included one of the weakest halos, NGC~4565
(Fig.~\ref{fig:n4565}) as well as one of the brightest, NGC~891
(Fig.~\ref{fig:n891}). Assuming energy density equipartition between
magnetic fields and cosmic rays, the scale height of the total
magnetic field is about 4~times larger than that of the synchrotron
emission, hence $\simeq7$~kpc on average. A prominent exception is
NGC~4631 with the largest radio halo (scale height of
$\simeq2.5$~kpc) observed so far \citep{hummel90,hummel91a,krause04}
(Fig.~\ref{fig:n4631cm22}). In case of energy equipartition, the
scale height of the total field is at least $(3+\alpha)$ times
larger than the synchrotron scale height (where $\alpha$ is the
synchrotron spectral index), hence $\ge10$~kpc in NGC~4631.

The scale height may be larger if cosmic rays originate from
star-forming regions in the plane and are not re-accelerated in the
halo, so that the electrons lose their energy above some height and
the equipartition formula yields too small values for the field
strength \citep{beckkrause05}. As the degree of linear polarization
increases with height above the disk midplane, the scale height of
the ordered field may be even larger.

\begin{figure}
\begin{center}
  \includegraphics[bb = 27 25 516 493,width=0.4\textwidth,clip=]{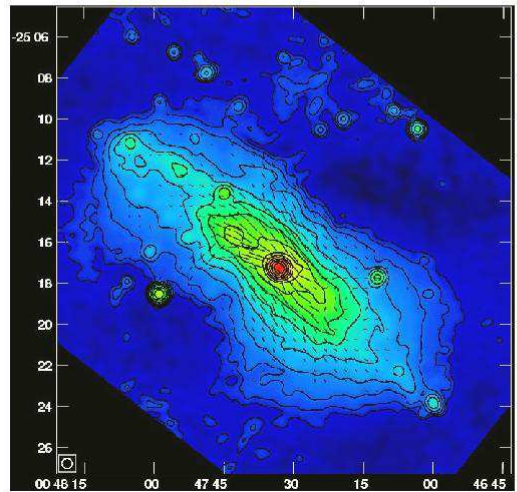}
 \end{center}
\caption{Total radio emission at 6~cm wavelength and polarization
vectors of the almost edge-on spiral galaxy NGC~253, combined from
observations with the VLA and the Effelsberg 100-m telescope
\citep{heesen05}. (Copyright: AIRUB Bochum / MPIfR Bonn) }
\label{fig:n253}
\end{figure}

Due to its large scale height, the magnetic energy density in halos
is much larger than that of the thermal gas \citep{ehle98}, while
still smaller than the dominating kinetic energy of the wind
outflow. Hence the magnetic field is frozen into the outflow and its
structure is a signature of the outflow kinematics
(Sect.~\ref{sect:halomf}).

Radio halos grow with decreasing observation frequency, which
indicates that the extent is limited by energy losses of the
cosmic-ray electrons, i.e. synchrotron, inverse Compton,
bremsstrahlung and adiabatic losses \citep{pohl90}. The stronger
magnetic field in the central regions leads to larger synchrotron
loss, leading to the squeezed shape of many halos, e.g. NGC~253
\citep{carilli92,beck94a,heesen05} (Fig.~\ref{fig:n253}), which is
in contrast to its almost spherical X-ray halo \citep{pietsch00}. On
the other hand, the exceptionally large and almost spherical radio
halo above the central region of NGC~4631 \citep{hummel90} (see also
Fig.~\ref{fig:n4631cm22}) indicates that the magnetic field is
weaker and/or the galactic wind is faster than in other galaxies.

The two different transport mechanisms for cosmic rays, diffusion or
advection (galactic wind), can be distinguished and modeled from the
variation of the radio spectral index with distance from the plane,
to be determined from multi-frequency radio observations
\citep{pohl90}. Breaks in the spectra of total radio emission due to
galactic winds were detected in several radio halos \citep{pohl91a}.
NGC~4631 is also exceptional concerning its radio spectrum
\citep{pohl91b}. \citet{heesen08} measured a transport speed of
about 200~km/s in the halo NGC~253 from the radio scale height and
the synchrotron loss time.

Note that most of the large-scale halo emission is missing in some
VLA maps \citep{sukumar91,carilli92,soida05}, which also severely
affects the polarization angles. A combination with single-dish
(Effelsberg) data is indispensable for large galaxies and at
wavelengths $\le6$~cm (Figs.~\ref{fig:n4565} and \ref{fig:n253}),
while at $\ge20$~cm the VLA is sensitive to extended structures up
to $\ge10\arcmin$, which is sufficient to detect the halos of the
largest edge-on galaxies (Fig.~\ref{fig:n4631cm22}).

\section{Magnetic field structure in halos}
\label{sect:halomf}

Radio polarization observations of nearby galaxies seen edge-on
generally show a disk-parallel field near the disk
plane \citep{dumke95}. High-sensitivity observations of several
edge-on galaxies like NGC~891 (Fig.~\ref{fig:n891}), NGC~5775
(Fig.~\ref{fig:n5775}), NGC~253 (Fig.~\ref{fig:n253}) and M~104
\citep{krause06} show vertical field components which increase with
increasing height $z$ above and below the galactic plane and also
with increasing radius, so-called X-shaped magnetic fields
\citep{krause07}.

The observation of X-shaped polarization patterns is of fundamental
importance to understand the origin of magnetic fields in halos. The
X-shape is inconsistent with the predictions from standard dynamo
models without wind flows (Sect.~\ref{sect:dynamo}), so that the
field has to be transported from the disk into the halo. A uniform
wind would just lift up the disk field and can also be excluded. A
superwind emerging from a starburst nucleus, as predicted for
NGC~253 \citep{heckman90}, would produce a radial pattern, which is
not observed. The most probable explanation is a wind emerging from
the disk which is modified by differential rotation in the halo and
dynamo action.

The analysis of the highly inclined galaxy NGC~253
(Fig.~\ref{fig:n253}) allowed a separation of the observed field
into an axisymmetric disk field and a halo field inclined by about
50\degr\ \citep{heesen08}. Similar tilt angles are also observed at
large heights in other edge-on galaxies.

The field lines in the outer halo of NGC~4631 suggest a dipole field
(Fig.~\ref{fig:n4631cm22}), but Faraday rotation measures do not
support this interpretation (see Sect.~\ref{sect:rm}). The radio
halo above the inner disk is composed of almost vertical magnetic
spurs connected to star-forming regions \citep{golla94}. The
observations support the idea of a strong ``spiky'' galactic wind
which is driven by regions of star formation in the inner disk
\citep{elstner95}. Furthermore, differential rotation is small in
the inner disk, so that vertical field lines are less twisted. At
larger radii, where star formation is weaker and differential
rotation is larger, the field becomes parallel to the disk. Another
galaxy with vertical fields away from the plane and hence a strong
wind is NGC~4666 \citep{dahlem97,soida05}.

\section{Faraday rotation and depolarization in halos}
\label{sect:rm}

Polarization ``vectors'' are ambiguous by 180\degr\ and hence cannot
distinguish between a halo field which is sheared into elongated
Parker loops or a regular halo field generated by a dynamo. A
large-scale regular field can be measured only by Faraday rotation
measures (RM). In external galaxies, the vertical field symmetry
could not yet been determined from existing RM data with sufficient
accuracy. RM values in halos, e.g. in NGC~253 \citep{beck94a} and in
NGC~4631, are small and do not show large-scale patterns. Indirect
evidence for preferred symmetric toroidal fields follows from the
possible dominance of the inward-directed radial field component in
the disk \citep{krause98}. Such a dominance of is in conflict with a
reversal of the toroidal field across the plane \footnote{Note that
the observed Faraday rotation measure RM is {\it not} zero along a
line of sight passing through a disk containing a field reversal in
its midplane if cosmic-ray electrons and thermal plasma are mixed in
the disk. In this case RM is half of that from a disk without a
reversal, and the sign of RM traces the field direction in the layer
which is {\it nearer} to the observer.} in which case the observed
field direction would depend on the aspect angle and no preference
would be expected for a galaxy sample.

In the Milky Way, the pattern of rotation measures from pulsars and
polarized background sources indicates a symmetric local magnetic
field, but an antisymmetric field in the halo, i.e. the toroidal
field reverses its sign across the plane (dynamo mode A0,
Fig.~\ref{fig:A0}) \citep{han97,han02,sun08}.

Faraday depolarization is another method to detect magnetic fields
and ionized gas in galaxy halos. In NGC~891 and NGC~4631 the mean
degree of polarization at 1.4~GHz increases from about 1\% in the
plane to about 20\% in the upper halo. This was modeled by
depolarization due to random magnetic fields of $10~\mu$G and
$7~\mu$G strength, respectively, and ionized gas with densities in
the plane of 0.03~cm$^{-3}$ and 0.07~cm$^{-3}$ and scale heights of
0.9~kpc and 1.3~kpc, respectively \citep{hummel91a}.

In face-on galaxies, structures in the halo may become visible via
depolarization of the underlying disk emission, e.g. in M~51
\citep{berk97}. A large depolarized region above the disk of
NGC~6946 may be due to a strong vertical field \citep{beck07a}.

\section{Interactions}

Interaction between galaxies or with a dense intergalactic medium
imprints unique signatures onto magnetic fields in galaxy halos and
thus onto the radio emission. The Virgo cluster is a location of
strong interaction effects. Highly asymmetric distributions of the
polarized emission shows that the magnetic fields of several
spirals are strongly compressed on one side of the galaxy
\citep{vollmer07,wez07}.

Interaction may also induce violent star-formation activity in the
nuclear region or in the disk which may produce huge radio lobes due
to outflowing gas and magnetic field. The lobes of the Virgo spiral
NGC~4569 reach out to at least 25~kpc from the disk and are highly
polarized \citep{chyzy06}. However, there is no indication for a
recent starburst, so that the radio lobes are probably a signature
of activity in the past.

Polarized radio emission is an excellent observational tracer of
interactions. As the decompression timescale of the field is very
long, it keeps memory of events in the past. These are still
observable if the lifetime of the illuminating cosmic-ray electrons
is sufficiently large. Radio observations at low frequencies are
preferable.

\section{Summary and Outlook}

Total and polarized radio emission of edge-on galaxies is a powerful
tool to study the physics of galaxy halos. The extent and spectral
index of radio halos gives information on the transport of
cosmic-ray electrons and their origin in star-forming regions in the
disk. Estimates of the electron lifetime allow an estimate of
the transport speed. The shape of the halo and the field
structure may allow to distinguish a global wind driven by star
formation in the disk from a superwind emerging from a central
starburst. The field structure in the halo reflects the interaction
between the outflow and differential rotation. Shock fronts between
colliding halo outflows may re-accelerate cosmic rays and enhance
the radio emission. Halo fields compressed by galaxy interactions or
ram pressure with the intergalactic gas are observable via polarized
emission. Faraday rotation measures trace regular fields and can
test dynamo models \citep{stepanov08}. Faraday rotation and
depolarization are also sensitive tools to detect ionized gas in
galaxies and in the intergalactic space.

Future radio telescopes will widen the range of observable phenomena
around galaxies. High-resolution observations at high frequencies
with the Extended Very Large Array (EVLA) and the planned Square
Kilometre Array (SKA) are required to show whether the halo fields
are regular or composed of many stretched loops. Forthcoming
low-frequency radio telescopes like the Low Frequency Array (LOFAR),
Murchison Widefield Array (MWA), Long Wavelength Array (LWA) and the
low-frequency SKA will be suitable instruments to search for
extended synchrotron radiation at the lowest possible levels in the
outer galaxy halos, the transition to intergalactic space and in
galaxy clusters and will give access to the so far unexplored domain
of weak magnetic fields in galaxy halos \citep{beck07b}. The
detection of radio emission from the intergalactic medium will allow
to probe the existence of magnetic fields in such rarified regions,
measure their intensity, and investigate their origin and their
relation to the structure formation in the early Universe.

As Faraday rotation angles increase with $\lambda^2$, low-frequency
telescopes will also be able to measure very small Faraday rotation
measures (RM) and Faraday depolarization, which allow to detect weak
magnetic fields and low-density ionized gas in galaxy halos. The
present limit in electron densities of about 0.003~cm$^{-3}$ can be
reduced by at least one order of magnitude.

The planned all-sky RM survey with the SKA at about 1~GHz will be
used to model the structure and strength of the magnetic fields in
the intergalactic medium, the interstellar medium of intervening
galaxies and in the Milky Way \citep{beck+04}. With deep
polarization and RM observations the origin and evolution of
galactic magnetic fields will be investigated. ``Cosmic magnetism''
has been accepted as Key Science Projects for both, LOFAR and SKA.

\end{document}